# In Cloud, Can Scientific Communities Benefit from the Economies of Scale?

Lei Wang, Jianfeng Zhan, Weisong Shi, *Senior Member IEEE*, and Yi Liang

**Abstract**—The basic idea behind Cloud computing is that resource providers offer elastic resources to end users. In this paper, we intend to answer one key question to the success of Cloud computing: in Cloud, can small or medium-scale scientific computing communities benefit from the economies of scale? Our research contributions are three-fold: first, we propose an enhanced scientific public cloud model (*ESP*) that encourages small or medium scale research organizations rent elastic resources from a public cloud provider; second, on a basis of the ESP model, we design and implement the *DawningCloud* system that can consolidate heterogeneous scientific workloads on a Cloud site; third, we propose an innovative emulation methodology and perform a comprehensive evaluation. We found that for two typical workloads: high throughput computing (HTC) and many task computing (MTC), *DawningCloud* saves the resource consumption maximally by 44.5% (HTC) and 72.6% (MTC) for service providers, and saves the total resource consumption maximally by 47.3% for a resource provider with respect to the previous two public Cloud solutions. To this end, we conclude that for typical workloads: HTC and MTC, *DawningCloud* can enable scientific communities to benefit from the economies of scale of public Clouds.

——————————— ◆ ———————————

## 1 INTRODUCTION

Traditionally, in scientific computing communities (in short, scientific communities), many small- or medium-scale organizations tend to purchase and build dedicated cluster systems to provide computing services for typical workloads. We call this usage model *the dedicated system model*. The dedicated system model prevails in scientific communities, of which an organization owns a small- or medium-scale dedicated cluster system and deploys specific *runtime environment software that is responsible for managing resources and its workloads*. A dedicated system is definitely worthwhile [35] as such a system is under the complete control of the principal investigators and can be devoted entirely to the needs of their experiment. However, there is a prominent shortcoming of the dedicated system model: for peak loads, a dedicated cluster system can not provide enough resources, while lots of resources are idle for light loads.

Recently, as resource providers [2], several pioneer computing companies are adopting the concept of infrastructure as a service, among which, Amazon EC2 contributed to popularizing the infrastructure-as-a-service paradigm [32]. A new term Cloud is used to describe this new computing paradigm [4] [17] [24] [32]. In this paper, we adopt the terminology from B. Sotomayor et al.'s paper [32] to describe different types of clouds. *Public clouds* offer a publicly accessible remote interface for *the masses*' creating and managing virtual machine instances within their proprietary infrastructure [32]. The primary aim of a private cloud deployment is to give local users a flexible and agile private infrastructure to manage workloads on their self-owning cloud sites [32]. Private cloud can also support *a hybrid cloud model* by supplementing local infrastructure with computing capacity from an external public cloud [32].

In scientific communities, more and more research groups feel great interest in utilizing open source cloud computing tools to build *private clouds*[23][32][36], or proposing *hybrid cloud models*[26][29][30] to augment their local computing resources with external public clouds. In this paper, we take a different perspective to focus on public clouds, and intend to answer the key question: *in public cloud, can small or medium scientific computing communities benefit from the economies of scale?* If the answer is yes, we can provide an optional cloud solution for scientific communities, which is complementary to state-of-art and state-of-practice private or hybrid cloud solutions, and hence many small and medium-scale scientific computing organizations can benefit from public clouds. Answering this question has two major challenges: first, cloud research communities need to propose cloud usage models and build systems that enable scientific communities to benefit from the economies of scale of public clouds; second, we need to present an innovative evaluation methodology to guide the design of *experiments to answer our concerned question,* since trace data of consolidating several scientific communities' workloads are not publicly available on production systems and experiments on large-scale production systems are also forbiddingly costly.

Previous efforts fail to resolve the above issues in several ways. First, while Armbrust et al. [2] in theory show a Web service workload can benefit from the economies of scale on a Cloud site, no one answers this

————————————————

- Lei Wang and Jianfeng Zhan are with Institute of Computing Technology, Chinese Academy of Sciences. E-mail: {wl, jfzhan}@ncic.ac.cn. Jianfeng Zhan is the corresponding author.
- Weisong Shi is with Department of Computer Science, Wayne State University. E-mail:weisong@cs.wayne.edu.
- Yi Liang is with Department of Computer Sciences, Beijing University of Technology.

question from the perspective of scientific communities. Second, in scientific communities, most of work focuses on private cloud or hybrid cloud solutions. For example, two open source projects, OpenNebula (www.opennebula.org) and Haizea1 (haizea. cs.uchicago.edu), are complementary and can be used to manage virtual infrastructures in private/hybrid clouds [32] for scientific communities. However, these research efforts help little in providing platforms for answering our concerned questions, since the concerns of private or hybrid clouds mainly revolve around activities (or workloads) of a single research institute or group. Third, state-of-art and state-of-practice public cloud solutions provide limited support for scientific communities. For example, some public cloud providers, such as Elastra and Rightscale, focus on deploying and managing *web services or database servers* on top of infrastructure-as-a-service clouds [32]; Amzaon's EC2 directly provides resources to end users, and relies upon end user's manual management of resources; Deelman et al. [7] propose that each staff of an organization (end users) directly leases virtual machine resources from EC2 in a specified period for running applications (which we call *Deelman's public cloud model or Deelman's model* in the rest of this paper). Evangelinos et al. [3] propose that an organization as a whole rents resources with the fixed size from EC2 to create a virtual cluster system that is deployed with a queuing system, like OpenPBS (which we call *Evangelinos's public cloud model or Evangelinos's model* in the rest of this paper). Our experiment results in Section 3 show that a) Deelman's solution will lead to high peak resource consumption, which raises challenge for the capacity planning of a system: b) Evangelinos's solution leads to high resource consumption because of its static resource management policy. Besides, little work supports consolidating heterogeneous scientific workloads on a cloud site. For example, in scientific communities, there are two typical workloads: <u>h</u>igh <u>t</u>hroughput <u>c</u>omputing (*HTC*) delivers large amounts of processing capacity over long period of time [1], and <u>m</u>any <u>t</u>ask <u>c</u>omputing (*MTC*) delivers much large numbers of computing resources over short period of time to accomplish many computational tasks [1]. Heterogeneous scientific workloads have different application characteristics and evaluation metrics, which we will explain in detail in Section 3.1, and hence have different requirements in terms of workload management and resource provisioning.

On the Dawning 5000 cluster system, ranked as top 10 of Top 500 super computers in November 2008 (http://www.top500.org/lists/2008/11), we design and implement an innovative system- *DawningCloud*. With DawningCloud, scientific communities (as *service providers*) do not need to own dedicated systems, and instead rent elastic resources from *a public cloud provider (as a resource provider)*. The contributions of our work can be concluded as follows:

First, we propose an innovative cloud usage model, called the <u>e</u>nhanced <u>s</u>cientific <u>p</u>ublic cloud model (ESP) for scientific communities. Different from *the dedicated system* and private/hybrid clouds, service providers in the ESP model do not need to own resources while *fully control* their runtime environments; unlike Deelman's and Evangelinos's public cloud models, service providers dynamically resize resources according to workload status.

Second, on a basis of the ESP model, we design and implement the *DawningCloud* system that can consolidate heterogeneous scientific workloads on one Cloud site. DawningCloud provides an enabling platform for answering our concerned question.

Third, we propose an emulation methodology to evaluate the system and conduct a comprehensive evaluation of DawningCloud and two public cloud solutions. For typical HTC and MTC workloads, our experiments show that: a) in comparison with the system incarnating Deelman's model, DawningCloud (which incarnates the ESP model) saves the resource consumption maximally by 44.5% (HTC) and 72.6% (MTC) for the service providers, and saves the total resource consumption by 44.7% for the resource provider ; b) in comparison with the system incarnating Evangelinos's model and a dedicated cluster system, DawningCloud saves the resource consumption maximally by 37.8% (HTC) and 67.5% (MTC) for the service providers, and saves the total resource consumption by 47.3% for the resource provider.

Forth, irrespective of specific workloads, we verify that DawningCloud can achieve the economies of scale on a Cloud platform using an analytical approach.

The organization of this paper is as follows: Section 2 presents the proposed ESP model; Section 3 presents the enabling system *DawningCloud*; Section 4 proposes an evaluation methodology and answers our concerned question in experiments; Section 5 verifies that DawningCloud indeed can achieve the economies of scale; Section 6 summarizes the related work; Section 7 draws a conclusion.

## 2 THE ESP MODEL

In this section, first we describe three roles in a Cloud site; second, we introduce the details of the ESP model; third, we list the distinguished features of the ESP model.

### 2.1 Three Players

We identify three roles in a Cloud site: *resource provider*, *computing service provider* and *end user*.

*A resource provider* owns a Cloud site (or a federated cloud systems [17]), and offers elastic resources to service providers in a pay-as-you-go manner [2].

Different from EC2, of which a resource provider directly offers resources to ends user, we identify another role: *computing service provider (in short, service provider). A service provider* acts as the proxy of an organization, leases resources from a resource provider and provides computing service to its end uses. Each staff in an organization plays the role of an *end user*. In this paper, we do not consider the case of which there are many compelling resource providers, so we presume that there are only one resource provider, several service providers and large amount of end users affiliated to each service provider in a typical Cloud site.

## 2.2 Details of the ESP Model

Fig.1 shows a typical scenario of the ESP model, of which two service providers rent resources from a public cloud provider, and consolidate their workloads on a Cloud site. In the rest of this section, we introduce the usage pattern of the ESP model.

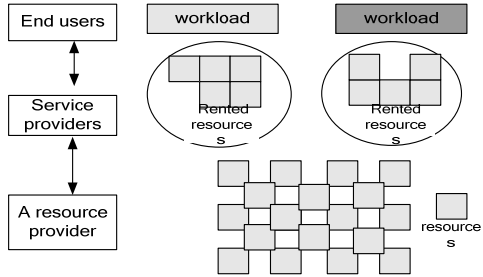

Fig. 1. A typical scenario of the ESP model.

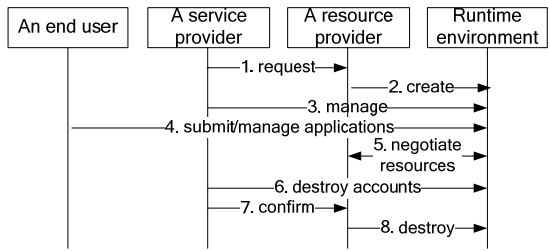

Fig. 2. The usage pattern of the ESP model.

As shown in Fig.2, the usage pattern is described as follows:

1) A service provider specifies its runtime environment requirements, including workload types: MTC or HTC, size of resources, types of operating system, and then requests a resource provider (which is a public cloud provider) to create a customized runtime environment. In our previous work [14], we have presented a runtime environment agreement for describing diverse runtime environment requirements of different service providers.

2) A resource provider creates a runtime environment for a service provider according to its requirement.

3) After a runtime environment is created, a service provider manages its runtime environment with full control, e.g. creating accounts for end users.

4) Each end user uses its accounts to submit and manage MTC or HTC applications in a runtime environment.

5) When a runtime environment is being providing services, a runtime environment can automatically negotiate resources with the proxy of a resource provider to resize resources by leasing more resources or releasing idle resources according to current workload status.

6) If a service provider wants to stop its service, it will inform its affiliated end users to backup data. Each end user can backup its data to storage servers provided by a resource provider. And then a service provider will destroy accounts of each end user in a runtime environment.

7) A service provider confirms a resource provider that the runtime environment is ready for destroying.

8) A resource provider destroys the specified runtime environment and withdraws the corresponding resources.

## 2.3 distinguished features of the ESP Model

Table 1 compares the ESP model with other models.

TABLE 1
THE COMPARISON OF DIFFERENT USAGE MODELS.

|  | DS | PC | HC | DP | EP | ESP |
|---|---|---|---|---|---|---|
| independent role of service provider | No | No | No | Yes | Yes | yes |
| service provider vs. resource provider | 1:1 | 1:1 | n:1 | n:1 | n:1 | n:1 |
| resources | local | local | local+ rented | rented | rented | rented |
| resource provisioning | fixed | fixed | fixed+ elastic | Fixed | fixed | elastic |

**DS** stands for dedicated system; **PC** stands for private cloud; **HC** stands for hybrid cloud; three public cloud models: **DP** stands for Deelman's public cloud; **EP** stands for Evangelinos's public cloud; **ESP** stands for our enhanced scientific public cloud model.

There are three distinguished features of the ESP model. First, our ESP model allows a resource provider to provision resources and provide runtime environments to $n$ $(n>>2)$ small or medium scale scientific communities, and hence it guides the design and implementation of the enabling platform helping us to answer the concerned economies of scale question. The dedicated systems and private clouds' limited use scopes will prevent service providers from benefiting from the economical of scale. In the private cloud model, only departments or groups belonging to the same organization share the same cloud resources. In hybrid clouds, users own local resources, and only request elastic resources from external public clouds for workload spike; besides, hybrid clouds that connect local resources with external public clouds are difficult for some parallel applications that rely heavily on frequent collective communications, since these applications are generally sensitive to network delays [30], and hence it may not benefit from using resources from multiple computing sites.

Second, the ESP model encourages the independent user role: *service provider*. In our model, a resource provider owns resources and creates runtime environment on demand for a service provider, while a service provider only rents resources and provides computing services to end users, and hence two user roles have *separated functions*. In private clouds, service providers are often affiliated with the resource provider, and the role of service provider is complexly intertwined with the role of resource provider. In the hybrid cloud model, users rent resources from the external public cloud as a service provider, at the same time they also own local resources as a resource provider.

Third, in the ESP model a service provider does not own resources, and instead automatically requests elastic resources from the resource provider according to workload status. In Deelman's public cloud model, each end user manually requests or releases resources from a resource provider. In Evangelinos's public cloud model, an organization as a whole obtains resources with the fixed size from a resource provider. In the dedicated system and private cloud models, in general they own fixed resources, though for the latter, a specific workload may use elastic resources within the organization for a specific duration. In hybrid clouds, users also rent elastic resources from the external public cloud; however, it is difficult for the others to share idle local resources when local loads are light.

## 3 AN ENABLING SYSTEM: DAWNINGCLOUD

To provide computing services, traditionally a small or medium scale organization owns a dedicated cluster system. Since different organization may have different work plans, their workloads may vary in the same period. We argue that on a Cloud site, the consolidation of large amount of heterogeneous scientific workloads may achieve the economies of scale. So, according to the ESP model and on a basis of our previous PhoenixCloud system [13] [14], we design and implement an enabling system, DawningCloud, for a resource provider to consolidate heterogeneous scientific workloads. In this paper, we mainly consider two workloads: HTC and MTC.

Our previous PhoenixCloud system has two major features: first, it presents a runtime environment agreement that expresses diverse runtime environment requirements of different service providers; second, it treats runtime environment as a first-class entity and enables creating coordinated runtime environments on demand. PhoenixCloud supports two workloads: web service applications and parallel batch jobs.

In this section, we introduce two most important features of DawningCloud: first, how to create a MTC or HTC runtime environment on demand on a Cloud site? Second, we propose automatic resource management mechanisms and policies for coexisting MTC or HTC runtime environments.

### 3.1 Requirement Differences of MTC and HTC.

Since there are diverse MTC workloads [1] and HTC workloads, in this paper, we take a typical MTC workload, Montage workflow (http://montage.ipac.acltech.edu), which is introduced by Ian Foster et al [1], and representative HTC workloads, batch jobs, which are presented in the condor project [11], to present the design of MTC and HTC runtime environments. In the DawningCloud design, we consider three requirement differences between MTC and HTC runtime environments as follows:
1) Usage scenes: the aim of HTC is designed for running parallel/sequential batch jobs; the aim of MTC is designed for running scientific workflows, like Montage workflow [1].
2) Application characteristics: MTC applications [1] can be decomposed to a set of small jobs with dependencies, whose running time is short; while batch jobs in HTC are independent and running times of jobs are varying.
3) Evaluation metrics: HTC service providers concern job's throughput over a long period of time; while MTC service providers concern job's throughput over a short period of time.

### 3.2 DawningCloud Architecture

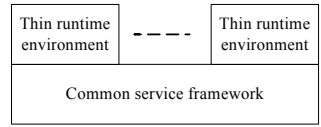

Fig. 3. The framework of DawningCloud.

As shown in Fig.3, we present a layered architecture for DawningCloud: one is the *common service framework* and the other is the *thin runtime environment*. The concept of thin runtime environment [14] indicates that the common sets of functions for different runtime environments are delegated to the common service framework, and a thin runtime environment only implements core functions for a specific workload.

The major functions of the common service framework are responsible for managing lifecycles of thin runtime environments, for example creating, destroying thin runtime environments, and provisioning resources to thin runtime environments in terms of nodes or virtual machines. The main services of the common service framework [14] are as follows:

*The resource provision service* is responsible for providing resources to different thin runtime environments.
*The lifecycle management service* is responsible for managing lifecycles of thin runtime environments.
*The deployment service* is a collection of services for deploying and booting operating system, the common service framework and thin runtime environments.
*The virtual machine provision service* is responsible for creating or destroying virtual machine like XEN.
*The agent* is responsible for downloading required software packages, starting or stopping service daemons.

In DawningCloud, on a basis of the common service framework, we implement two kinds of thin runtime environments: *MTC thin runtime environment* and *HTC thin runtime environment*.

In HTC thin runtime environment, we only implement three services: *the HTC scheduler*, *the HTC server* and *the HTC web portal*. The HTC scheduler is responsible for scheduling users' jobs through a *scheduling policy*. *The HTC server* is responsible for dealing with users' requests, managing resources, loading jobs. *The HTC web portal* is a GUI through which end users submit and monitor HTC applications.

In MTC thin runtime environment, we implement four services: *the MTC scheduler*, *the MTC server*, *the trigger monitor* and *the MTC web portal*. The function of *the MTC scheduler* is similar to *the HTC scheduler*. Different from the HTC server, *the MTC server* needs to parse a workflow description model, which are inputted by users on *the MTC web portal*, and then submit a set of jobs/tasks with dependencies to *the MTC scheduler* for scheduling. Besides, a new service, *the trigger monitor*, is responsible for monitoring trigger conditions of a workflow, such as changes of database's records or files, and notifying changes to *the MTC server* to drive running of jobs in different stages of a workflow. *The MTC web portal* is also much more complex than that of HTC, since it needs to provide a visual editing tool for end users to draw different workflows.

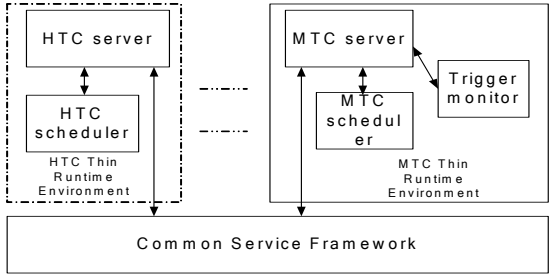

Fig.4. Coexisting MTC and HTC runtime environments on a basis of the common service framework.

Fig.4 shows a typical DawningCloud system, of which a MTC thin runtime environment and a HTC thin runtime environment reuse the common service framework.

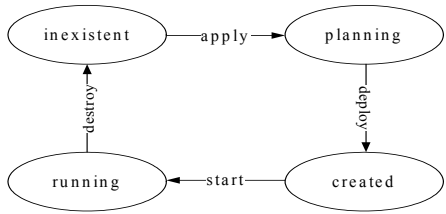

Fig.5. The lifecycle of a thin runtime environment.

As shown in Fig.5, the lifecycle of a thin runtime environment includes four main states: *inexistent, planning, created and running*. For a typical thin runtime environment, the lifecycle is as follows:
1) The initial state of a thin runtime environment is *inexistent*. The common service framework is running on a Cloud site. A service provider uses *the web portal of the common service framework* to *apply* for a new thin runtime environment. *The web portal of the common service framework* sends a request of creating a runtime environment to *the lifecycle management service* of the common service framework.
2) *The lifecycle management service* validates the request. If the request is valid, it marks the state of a new thin runtime environment as *planning*.
3) *The lifecycle management service* sends a request of *deploying a thin runtime environment* to agents on each related node, which then request *the deployment service* to download required software packages of a specific thin runtime environment. After a new thin runtime environment is deployed, *the lifecycle management service* marks its state as *created*.
4) *The lifecycle management service* sends the configuration information of the new thin runtime environment to *the resource provision service*.
5) *The lifecycle management service* sends a request to agents that *start* each component of a new thin runtime environment according to their dependencies. When components are started, command parameters will tell them what configuration parameters should be read. Then *the lifecycle management service* marks the state of the new thin runtime environment as *running*.
6) The new thin runtime environment begins providing services to end users. End users use the web portal to submit their applications.
7) According to current load status, the thin runtime environment dynamically requests or releases resources from or to *the resource provision service*.
8) If a service provider uses *the web portal of common service framework* to *destroy* its thin runtime environment, *the web portal of the common service framework* sends a request of destroying a thin runtime environment to *the lifecycle management service*; *the lifecycle management service* validates the information and destroys a thin runtime environment through prompting end users to backup data, stopping related daemons and offloading related software packages.

### 3.3 Dynamic Resource Negotiation Mechanism

We present the dynamic resource negotiation mechanism in DawningCloud as follows:
1) A service provider specifies its requirement for resource management in *a resource management policy*. A resource management policy defines the behavior specification of the HTC or MTC server in that the server resizes resources to what an extent according to what criterion. According to a resource management policy, the MTC or HTC server decides whether and to what an extent resizes resources according to current workload status, and then sends requests of obtaining or releasing resources to the resource provision service.
2) A resource provider specifies its requirement for resource provisioning in *a resource provision policy*, which determines when the resource provision service provisions how many resources to different thin runtime environments in what priority. According to a resource provision policy, the resource provision service decides to assign or reclaim how many resources to or from a thin runtime environment.
3) *A setup policy* determines when and how to do the setup work, such as wiping off the operating system or doing nothing. For each time of node assignment or reclaiming, a setup policy is triggered, and the lifecycle management service is responsible for performing the setup work.

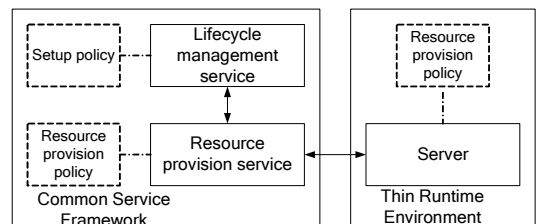

Fig. 6. Dynamic resource negotiation mechanism.

## 3.4 Resource Management and Provisioning Policies

In this section, we respectively propose resource management and provisioning policies for MTC and HTC service providers.

**Resource Management Policies:** We propose a resource management policy for a HTC or MTC service provider as follows:

There are two types of resources that are provisioned to a runtime environment: *initial resources* and *dynamic resources*. Once allocated to a HTC or MTC thin runtime environment, *initial resources* will not be reclaimed by the resource provision services until the thin runtime environment is destroyed. On the contrary, *dynamic resources* assigned to a thin runtime environment may be reclaimed by the resource provision service when a thin runtime environment is in the state of *running*.

In DawningCloud, a service provider and a resource provider needs to set four types of parameters in the resource management policy: a) *the size of initial resources*; b) *the time unit of leasing resources. A lease term of a dynamic resource must be the time unit of leasing resources times an integer.* For example, in EC2, the time unit of leasing resources is one hour; c) the *checking resource cycle*. It is a periodical timer that the *HTC or MTC server* checks jobs in queue; d) the *threshold ratio of obtaining dynamic resources*. For a thin runtime environment, a service provider needs to set a *threshold ratio of obtaining dynamic resources*, according to which the HTC or MTC server decides to whether to request more dynamic resource and request how many dynamic resources if requesting more resources is needed.

We propose a resource management policy for a HTC or MTC service provider as follows:

1) At the startup of a runtime environment, a service provider will request *initial resources* with the specified size.

2) The *HTC or MTC server* scans jobs in queue per *checking resource cycle*. If *the ratio of obtaining dynamic resources* exceeds *the threshold ratio or* the ratio of the *resource demand of the present biggest job in queue* to *the current resources owned by a thin runtime environment is greater than one* (which indicates that if the server does not request more resources, the present biggest job may not have enough resources for running), the server will request dynamic resources with *the size of DR* as follows:

*DR=the accumulated resources demand of all jobs in the queue – the current resources owned by the thin runtime environment.*

In this policy, we define *the ratio of obtaining dynamic resources* as the ratio of *the accumulated resource demands of all jobs in the queue* to *the current resources owned by a thin runtime environment*.

3) After obtaining dynamic resources from *the resource provision service*, the server registers *a new periodical timer* and checks idle dynamic resources per *time unit of leasing resources*. If there are idle dynamic resources with the size that is less than the value of DR, the server will release resources with the size of *DR= (DR- idle dynamic resources)*; else if there are idle dynamic resources with the size that is *equal to* or *more than the value of DR*, the server will release resources with the size of the DR and *deregisters the timer*.

There is only one difference in two resource management policies respectively proposed for a MTC or a HTC service provider: we need to set the *checking resource cycle* of MTC as a smaller value than that of HTC, this is because MTC tasks often run over in seconds and HTC jobs often run over in a longer period. We will discuss parameter configurations in Section 4.6.

**Resource provisioning policy:** Since our aim is to consolidate workloads of small or medium scale organizations on a Cloud site, we presume that in public clouds, a resource provider owns enough resources that can satisfy resource requests of *N* HTC and MTC service providers ($N>>2$). So we propose a simple resource provision policy for a service provider as follows:

First, *the resource provision service* provisions the requested initial resources to a thin runtime environment at its startup.

Second, when *the server* of a thin runtime environment requests dynamic resources, *the resource provision service* assigns enough resources to the server.

Third, when the server of a thin runtime environment releases dynamic resources, the resource provision service will passively reclaim resources released by the server.

## 4. EVALUATION METHODOLOGY AND EXPERIMENTS

In this section, first, we report our chosen workloads; second, we present the evaluation mythology; third, we give out the experiment configurations, and finally we will compare DawningCloud with the other three systems.

### 4.1 Workloads

We choose two typical HTC workload traces from the Parallel Workloads Archive: http://www.cs.huji.ac.il/labs/parallel/workload/. The utilization rate of all traces in the Parallel Workloads Archive varies from 24.4% to 86.5%. We choose one trace with lower load-NASA iPSC trace (46.6% utilization) and one trace with higher load-SDSC BLUE trace (76.2% utilization). The scales of NASA trace and BLUE trace are respectively 128 and 144 nodes, which are popular in small or medium organizations.

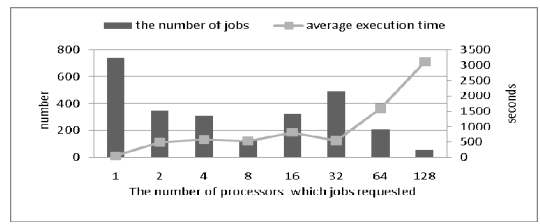

Fig. 7. Two weeks' NASA trace.

We randomly extract two weeks' traces of *the NASA iPSC trace* from Fri Oct 01 00:00:03 PDT 1993, and *the SDSC BLUE trace* from Apr 25 15:00:03 PDT 2000. For NASA trace, the average execution time is 575 seconds, and the total number of jobs is 2604, of which the distribution of resource demands of jobs in terms of processors and average execution time vs. different resource demands of jobs are shown in Fig.7. For BLUE trace, the average execution time is 2092 seconds, and the total number of jobs is 2666, of which the distribution of resource demands of jobs in terms of processors and average execution time vs. different resource demands of jobs are shown in Fig.8.

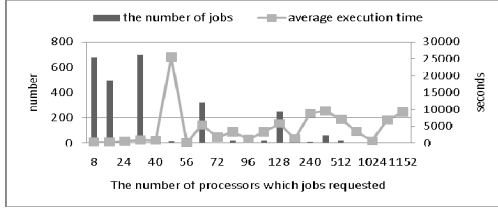

Fig. 8. Two weeks' BLUE trace.

For MTC, we choose a typical workload, Montage workflow (http://montage.ipac.caltech.edu). Montage is an astronomy workflow application, created by NASA/IPAC Infrared Science Archive for gathering multiple input images to create custom mosaics of the sky. The workload generator can be found on the web site http://vtcpc.isi.edu/pegasus/index.php/WorkflowGenerator, and the workload file includes the job name, run time, inputs, outputs and the list of control-flow dependencies of each job. The chosen Montage workload includes 9 types of task with the total amount of 1,000 jobs. Each job requests one node for running, and the average execution time of jobs is 11.38 seconds. The number of jobs and average execution time corresponding with each kind of task are shown in Fig.9. The run time of the MTC workload is shorter than that of HTC workload traces. In order to achieve the same duration (two weeks) like that of HTC workloads, we synthesize the MTC workload through repeatedly submitting Montage workload.

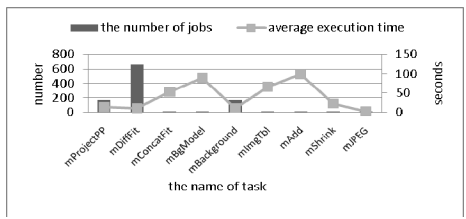

Fig. 9. The Montage workload trace.

### 4.2 Evaluation Metrics

In our experiments, we mainly concern the public cloud solutions, and hence a resource provider respectively chooses *DawningCloud*, the system incarnating Deelman's model (in short, *Deelman's system*), the system incarnating Evangelinos's model (in short, *Evangelinos's system*) and the dedicated system to provide computing services. For DawningCloud, Evangelinos's system and Deelman's systems, one resource provider owns a cloud platform.

We choose the number of completed jobs [10] to evaluate the performance metric of HTC service providers; and we choose the number of tasks per second [1] to reflect the performance metric of a MTC service provider. For a service provider, we choose the resource consumption in terms of node*hour to evaluate its cost. That is to say, for two weeks' NASA iPSC trace or SDSC BLUE trace, we sum the product of the consumed resources in terms of nodes and their corresponding consumed hours as the cost of a HTC service provider. In the Deelman's system, there is no role of a service provider, so we calculate the accumulated resource consumption of all end users, which amounts to the cost of a service provider in other models. For the dedicated cluster system, since a service provider owns resources, we calculate the resource consumption of a service provider as the product of the configuration size of a dedicated cluster system and the duration of a certain period.

For a resource provider, we choose *the total resource consumption* in terms of *node*hour* to evaluate the cost, which is the sum of all service providers' *resource consumptions*. Especially, we care about the *peak resource consumption* in terms of *nodes* in a certain period. *For the same workload, if the peak resource consumption of a system is higher, the capacity planning of a system is more difficult.*

In DawningCloud, Deelman's and Evangelinos's systems, since allocating or reclaiming resources will trigger setup actions, and we use *the accumulated times of adjusting nodes*, which is the sum of all service providers' *times of adjusting nodes*, to evaluate the *management overhead* of a resource provider.

The above metrics are obtained in the same period that is just the duration of workload traces (two weeks).

### 4.3 Evaluation Methodology

In the rest of this section, most of experiments are done with the emulation methodology, while we obtain the overhead of adjusting a node on the real system. Choosing the emulation methodology is based the following observations:

1) To evaluate a system, many key factors have effects on experiment results, and we need to do many times of time-consuming experiments, since durations of workload traces are several weeks. Through using an emulation method, we can speedup experiments and complete large amount of experiments within the shorter period of time.

2) With the real systems, consolidating several scientific communities' workloads needs hundreds of nodes, which results in mass resource requirements. While through using the emulation method, we can eliminate this resources limitation.

In this paper, all of the emulation systems are deployed on a test bed composed of nodes with the configuration of two AMD Opteron CPU, 2G memories and CentOS 5.0 operating system.

For each emulated system, *the job simulator* is used to emulate the process of submitting jobs. For HTC workload, the job simulator generates each job by extracting its submission time, real run time and requested number of nodes from the workload trace file; For MTC workload, the job simulator reads the workflow file, which includes submission time, real run time, requested number of nodes and dependencies between each job, and then submits jobs according to dependency constraints. We speed up the submission and completion of jobs by a factor of 1000.

In the rest of this section, we introduce how to emulate systems for different models.

**The emulated dedicated cluster systems**: For each dedicated cluster system, we deploy the *simplified DawningCloud* with two simulation modules: *the resource simulator* and *the job simulator on the testbed*. The resource simulator defines the configurations of the dedicated cluster system. Since the workload files are obtained from platforms with different configurations. For example, the NASA trace is obtained from a cluster system with each node composed of one CPU; and BLUE trace is obtained from a cluster system with each node composed of eight CPUs. In the rest of this paper, we presume that *each node in our simulated cluster is composed of one CPU*. And then, we scale workload traces with different constant values to the same configuration of the simulated cluster. Besides, the resource simulator not only simulates resource requesting and releasing, but also job managing such as loading or killing jobs and so on. The resource limit is enforced by the resource simulator. For HTC runtime environment, the job simulator reads the job information from workload trace file and submits the job to the HTC server; for MTC runtime environment, the job simulator replaces the trigger monitor to read the job information from workload trace file and analyze control-flow dependencies among jobs to decide submitting the right job to the MTC server.

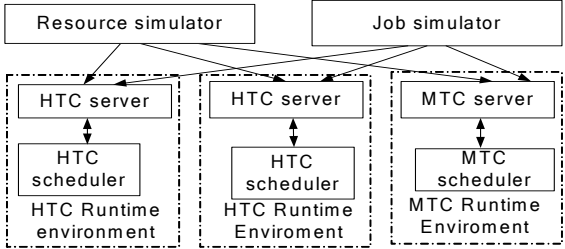

Fig.10. three emulated dedicated cluster systems for HTC and MTC.

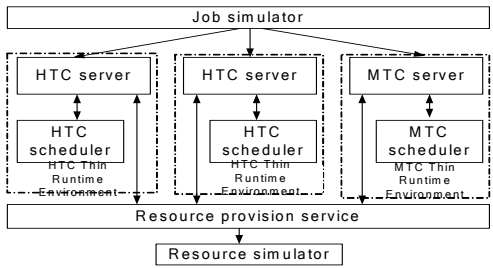

Fig.11. The emulated DawningCloud.

**Emulated DawningCloud system:** As shown in Fig.11, we deploy the *simplified* DawningCloud system, which keeps the resource provision service, one server and one scheduler for each thin runtime environment while removing other services. Resource requesting and releasing are simulated by the interactions between the resource provision service and the resource simulator.

**Emulated Deelman's system:** Since in Deelman's model, end users in scientific communities directly use EC2 for scientific computing. Based on the framework of DawningCloud, we implement and deploy an EC2-like system as shown in Fig.12 on the testbed with two simulation modules: the job simulator and the resource simulator. With respect to the real DawningCloud system, we only keep the resource provision service and the VM provision service. Resource requesting and releasing are enforced by the interactions between the resource provision service and the resource simulators. VM creating and destroying are enforced by the interactions between the VM provision service and the resource simulators. The job simulator reads the number of nodes which each job requests in the trace file and sends the request to the resource provision service, which assigns corresponding resources for each job. When each job runs over, the job simulator will release resources to the resource provision service.

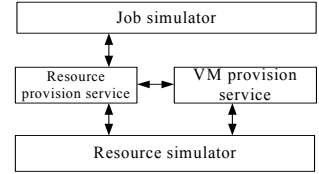

Fig.12 The emulated Deelman's system

**Emulated Evangelinos's system:** We implement the Evangelinos's system based on the framework of DawningCloud. Since a service provider in Evangelinos's system leases resources with the fixed size in a period from a resource provider at a time, so the emulated Evangelinos's system is closely similar to that of the dedicated cluster system, shown in Fig.10.

### 4.4 Experiment Configurations

In the following experiments, we emulate a public cloud scenario in which there are only one resource provider, two organizations providing HTC services and one organization providing MTC service. Of course, with our methodology, we can easily extend to the case that one resource provider provisions resources to more service providers.

**Resource configurations:** Since the maximal resource requirements of the NASA and BLUE traces are respectively 128 and 144 nodes, we respectively set the configuration sizes of the dedicated cluster systems for NASA trace and BLUE trace as 144 nodes and 128 nodes in the emulation experiments. For the Montage workload, because *the accumulated resource demand of all jobs in the queue* in most of running time is 166 nodes, we set the configuration size of the dedicated cluster system for Montage workload as 166 nodes to improve throughput in terms of *tasks per second* in the emulation experiments. In the emulated Evangelinos's system, the fixed lease term of resources is two weeks. Same like that of three dedicated cluster systems, *the sizes of leased resources* are respectively 128, 144 and 166 nodes for NASA, BLUE and Montage. DawningCloud and Deelman's system request elastic resources according to current workload traces.

**Scheduling policies:** A scheduling policy is needed by schedulers in DawningCloud, Evangelinos's system and the dedicated cluster system. In this paper, *we do not investigate the effect of different scheduling policies,* so we simply choose the *first fit scheduling policy* for HTC. The first-fit scheduling algorithm scans all the queued jobs in the order of job arrival and chooses the first job, whose resources requirement can be met by the system, to execute. For MTC workload, firstly we generate the job flow according to dependency constraints, and then we choose the FCFS (First Come First Served) scheduling policy in DawningCloud, Evangelinos's system and dedicated cluster systems. We respectively set *the scheduling cycle* of the HTC and MTC schedulers as 60 seconds and 1 second. The Deelman's system uses no scheduling policy, since all jobs run immediately without queuing.

**Resource management and provisioning policies:** DawningCloud, Deelman's and Evangelinos's systems adopt the same resource provisioning policy stated in Section 4.4. For the dedicated cluster systems, they own static resources. DawningCloud adopts the resource management policy proposed in Section 3.4, while the dedicated cluster system and Evangelinos' systems adopts the static resource management policy. Just like EC2, Deelman's system relies on the manual resource management.

### 4.5 System-level Evaluation

In DawningCloud, we need to set the following parameters for the service provider:

*a) The time unit of leasing resources (which is represented as C).* Time unit of leasing resources has effect on both DawningCloud and Deelman's system. When the time unit of leasing resources is shorter, resources will be adjusted more frequently, which brings higher management overhead.

*b) The size of initial resources (*which is represented as *B).*

*c) The checking resource cycle (*which is represented as S*).*

We set S as the same value of the scheduling cycle in the scheduling policy.

*d) The threshold ratio of obtaining dynamic resources* (which is represented as *R*).

Before reporting experiment results, we pick the following parameters as the baseline for comparison, and detailed parameter analysis will be deferred to Section 4.6.

Through comparisons with large amount of experiments, we set the baseline configurations in DawningCloud: [60C/40B/1.5R/60S] for HTC workload and [60C/20B/8R/1S] for MTC workload, where [60C] indicates that *the time unit of leasing resources* is 60 minutes.

TABLE 2
THE METRICS OF THE SERVICE PROVIDER FOR NASA TRACE

| Configuration | number of completed jobs | resource consumption (node*hour) | saved resources |
|---|---|---|---|
| dedicated cluster system | 2603 | 43008 | / |
| Evangelinos's system | 2603 | 43008 | 0 |
| Deelman's system | 2603 | 52943 | -23.1% |
| DawningCloud | 2603 | 29373 | 31.7% |

For the dedicated cluster system and Evangelinos's system, they have the same configurations with the only difference in that a service provider in the dedicated cluster system owns resources while a service provider in the Evangelinos's system leases resources, so they gain the same performance.

TABLE 3
THE METRICS OF THE SERVICE PROVIDER FOR BLUE TRACE

| Configuration | number of completed jobs | resource consumption (node*hour) | saved resources |
|---|---|---|---|
| dedicated cluster system | 2649 | 48384 | / |
| Evangelinos's system | 2649 | 48384 | 0 |
| Deelman's system | 2657 | 35838 | 25.9% |
| DawningCloud | 2657 | 30100 | 37.8% |

Table 2, Table 3 and Table 4 summarize the experiment results of two HTC service providers and one MTC service providers with DawningCloud, Evangelinos's system, dedicated cluster system and Deelman's systems. *The percent of the saved resources* are obtained against the resource consumption of the dedicated cluster system.

TABLE 4
THE METRICS OF THE SERVICE PROVIDER FOR MONTAGE WORKFLOW

| Configuration | tasks per second | resource consumption (node*hour) | saved resources |
|---|---|---|---|
| dedicated cluster system | 2.46 | 55776 | / |
| Evangelinos's system | 2.46 | 55776 | 0 |
| Deelman's system | 2.68 | 66200 | -18.7% |
| DawningCloud | 2.46 | 18108 | 67.5% |

For NASA trace and BLUE trace, in comparison with the dedicated cluster system/Evangelinos's systems, service providers in DawningCloud save the resource consumption maximally by 37.8% and minimally 31.7%, and at the same time gain the same or higher throughputs. This is because service providers in DawningCloud can resize resources according to workload status, while service providers in the dedicated cluster system/Evangelinos's systems owns or leases resources with the fixed size.

For Montage workload, DawningCloud has the same performance as that of the dedicated cluster system/Evangelinos's systems for the service provider. This is because driven by the resource management policy stated in Section 3.4, the MTC server will adjust dynamic resources to *the size of the accumulated resource demand of jobs in queue*, which is same as the chosen configurations of the dedicated cluster system/Evangelinos's systems (166 nodes). In comparison with the dedicated cluster system/Evangelinos's systems, the service provider in the DawningCloud saves the resource consumption by 67.5%, this is because the service provider in the dedicated cluster system/Evangelinos's systems owns or leases resources with the fixed size, while the service provider in DawningCloud owns *initial resources* with the smaller size, and resizes dynamic resources driven by the change of workload status.

For NASA trace and BLUE trace, with respect to the Deelman's system, DawningCloud saves the resource consumption maximally by 44.5% for service providers with the same performance. This is because the dynamic resource negotiation and queuing based resource sharing mechanisms in DawningCloud lead to the decrease of resource consumption. On the other hand, in Deelman's model, each end user directly obtains resources from the resource provider, which results in that Deelman's system consumes more resources than that of DawningCloud.

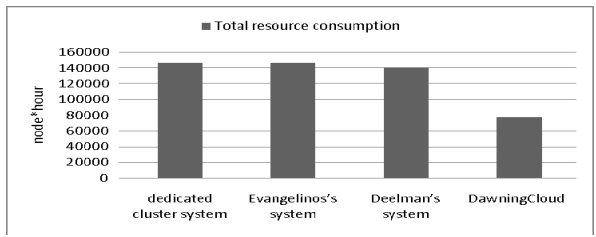

**Fig.13. Total resource consumption of the resource provider using four different systems.**

For Montage workload, DawningCloud saves the resource consumption by 72.6% with respect to that of the Deelman's system for the same service provider. This is because the required resources of end users will be provisioned immediately in the Deelman's system and the peak resource demand of MTC workload is high. At the same time, the Deelman's system gains higher throughput than that of DawningCloud.

Fig.13 and Fig.14 show experiment results for the resource provider using four different systems: DawningCloud, Evangelinos's system, Deelman's system and the dedicated cluster systems.

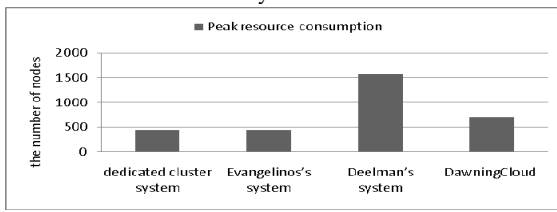

Fig.14. Peak resource consumption of the resource provider using four different systems.

Using the dedicated cluster system and Evangelinos's systems, the resource provider has the same total resource consumption and the same peak resource consumption, since they have only one difference in that the former owns resources while the latter leases resources.

Using DawningCloud, the total resource consumption of the resource provider is 77581 node*hour, which saves the total resource consumption by 47.3% with respect to that of the dedicated cluster system/Evangelinos's systems. In the dedicated cluster system/Evangelinos's systems, the service providers lease or purchase resources with *the fixed size that is decided by the peak resource demand of the largest job*. In contrast, in DawningCloud, a service provider can start with the small-sized initial resources and resize dynamic resources according to varying resource demand. Hence, the total resource consumption of DawningCloud is less than that of the dedicated cluster system/Evangelinos's systems when workloads of three service providers are consolidated. At the same time, With DawningCloud, the peak resource consumption of the resource provider is 705 nodes, which is only 1.61 times of that of dedicated cluster system/Evangelinos's systems.

Using DawningCloud, the resource provider saves the total resource consumption by 44.7% with respect to that of the Deelman's system, and the peak resource consumption of DawningCloud is only 0.45 times of that of the Deelman's system. Because for each job, the required resources will be provisioned immediately in the Deelman's system, its peak resource consumption is larger than that of DawningCloud.

Fig.15 shows the management overhead of the resource provider using Evangelinos's model, Deelman's model and DawningCloud. For dedicated cluster system, since the resource provider owns resource, it has no management overhead in terms of obtaining dynamic resources. From Fig.15, we can observe that the Evangelinos's system has the lowest management overhead, since it leases resources with the fixed duration. DawningCloud has smaller management overhead than that of the Deelman's system, since the initial resources will not be reclaimed until a runtime environment is destroyed.

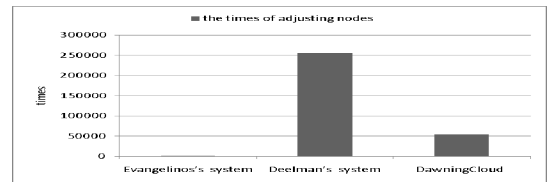

Fig.15 The management overhead of the resource provider.

In our real test, excluding wiping off the operating system, the total cost of assigning and reclaiming one node is 15.743 seconds, which includes the operation of stopping and uninstalling previous runtime environment packets, installing and starting new runtime environment packets. That is to say that the resource consumption of adjusting nodes in DawningCloud is approximately 120.6 node*hour when the total resource consumption is 77581 node*hour. This overhead is acceptable.

### 4.6 Parameter Analysis

Because of space limitation, we are unable to present the data for the effect of all parameters; instead, we constrain most of our discussion to the case that one or two parameters varies while the other parameters keep the same as those of the baseline configuration in Section 4.5, which are representative of the trends that we observe across all cases.

**The effect of the size of initial resources and the threshold ratio of obtaining dynamic resources.**

To save space, in DawningCloud we tune the size of initial resources (*B*) and the threshold ratio of obtaining dynamic resources (*R*) at the same time, while other parameters are [60C/60S] for HTC workload and [60C/1S] for MTC workload.

We respectively set *B* as (0,20,40,60,80,100,144) for BLUE workload and (0,20,40,60,80,100,128) for NASA workload, and (0,20,40,60,80,100,166) for Montage workload; at the same time, we tune *R* as (1,1.2,1.5,2,4,100) for HTC workloads and (1,2,4,8,16,100) for MTC workload.

Fig.16 shows the effect of different parameters. In Fig.16, *B0_R1* indicates that *B* is 0 and *R* is 1.

For three workload traces, when *the size of initial resources is the same like that of the Evangelinos's system and the threshold ratio of obtaining dynamic resources is so high as to no dynamic resource will be obtained*, DawningCloud has the same performance metrics as that of the Evangelinos's system. For example, for NASA, BLUE and Montage workload, the configuration is respectively B128_R100, B144_R100, and B166_R100.

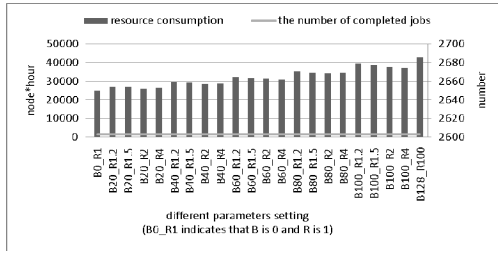

Fig.16-1. Resource consumption and the number of completed jobs V.S. different parameters setting for NASA trace.

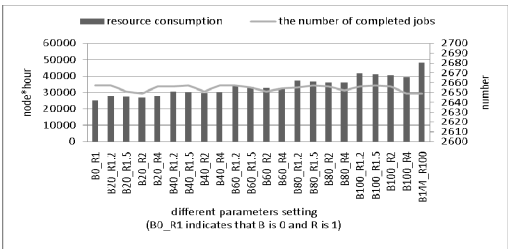

Fig.16-2. Resource consumption and the number of completed jobs V.S. different parameters setting for BLUE trace.

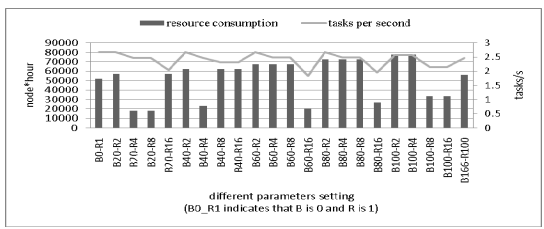

Fig.16-3. Resource consumption and tasks per second VS. different parameters for Montage workload.

For HTC workloads, we have the following observations:

1) In DawningCloud, the resource consumption is proportional to the size of initial resources; this is because initial resources are statically allocated to a service provider in DawningCloud. For the same workload, the size of initial resources increases, idle resources will also increases. When the size of initial resources is below the configuration size of Evangelinos's model, the size of initial resources has no significant effect on the number of completed jobs, since dynamic resources can be obtained in DawningCloud.

2) The resource consumption is inversely proportional to *R* when dynamic resources are not zero; this is because that larger threshold ratio can result in less opportunity of obtaining dynamic resources. There is no obvious relationship between R and the number of completed jobs.

For MTC workloads, we have the following observations:

1) There are several configurations (such as *B20_R4*, *B20_R8* and *B40_R4* in Fig.13) that make the service provider consumes less resources. From our observation, we found those configurations satisfying the empirical formulas ($B*R < RA$) and ($RA*R > RM$), where *RA* is *the accumulated resource demand of jobs in queue* in most of running time, and *RM* is the maximal accumulated resource demand of jobs in queue in running time. For Mantage workload trace, *RM* is 662, and *RA* is 166.

2) There is no obvious relationship between B and the resource consumption or the number of tasks per second.

3) There is no obvious relationship between R and the resource consumption or the number of tasks per second.

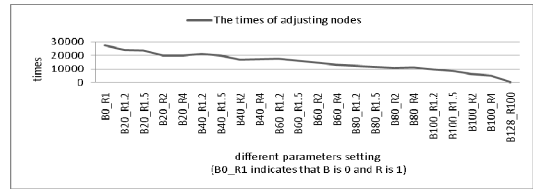

Fig.17-1. the service providers' times of adjusting nodes in two weeks VS. different parameters setting for NASA trace.

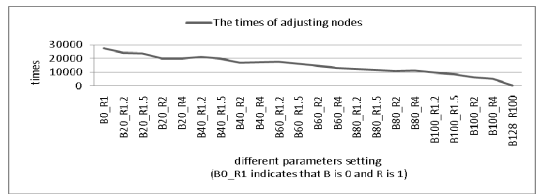

Fig.17-2. the service providers' times of adjusting nodes in two weeks VS. different parameters setting for BLUE trace.

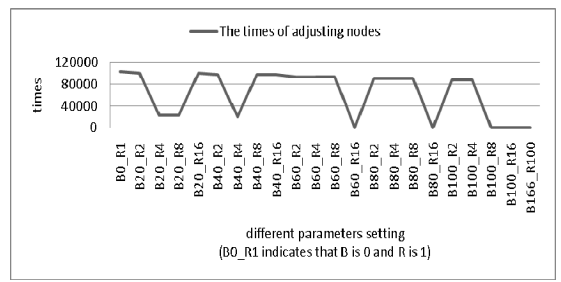

Fig.17-3. the service providers' times of adjusting nodes in two weeks VS. different parameters setting for Montage workload.

Fig.17 shows the effects of *the size of initial resource* (B) and *the threshold ratio of obtaining dynamic resources* (R) on the management overhead of the resource provider, which is the sum of all service providers' times of adjusting nodes.

For HTC workloads, we have the following

observations:

1) The management overhead of the service provider is inversely proportional to B; this is because more initial resource, less times of obtaining dynamic resources.

2) The management overhead is inversely proportional to R; this is because larger threshold ratio, less times of obtaining dynamic resource.

For MTC workloads, we have the following observations:

1) For some configurations satisfying the empirical formulas *(B\*R < RA) and (RA\*R > RM)* or *(B\*R > RM)*, the management overhead is less than that of other configurations.

2) There is no obvious relationship between the times of adjusting nodes for the MTC service provider and B/R.

**The effect of checking resource cycle.**

In DawningCloud, we respectively set *the checking resource cycle S* as 10/30/60/100/120/150/180/200 seconds, while other parameters are [60C/40B/1.5R] for HTC workloads. In the scheduling policy, *the scheduling cycle* is the same amount as *S*.

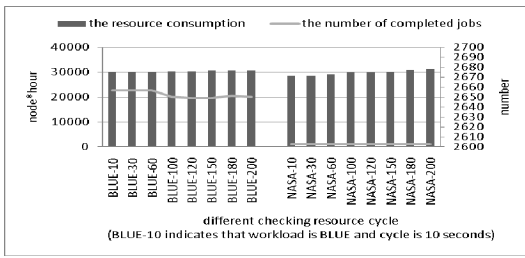

Fig.18-1. resource consumption and the number of completed jobs VS. checking resource cycle for BLUE and NASA trace.

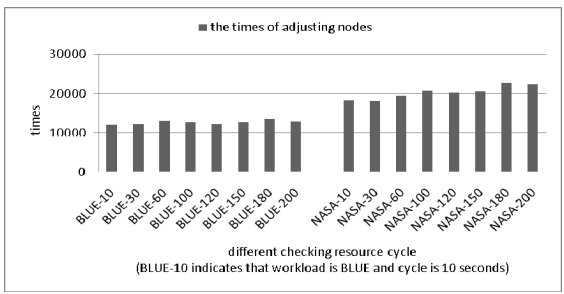

Fig.18-2. the service providers' times of adjusting nodes VS. checking resource cycle for BLUE and NASA trace.

In Fig.18, BLUE-10 indicates that the workload is BLUE and *the checking resource cycle* is 10 seconds. From Fig.18, we can observe that: *S* has small impact on the resource consumption, the number of completed jobs and the service provider's times of adjusting nodes. So we just set S as the same value of the scheduling cycle. Here is 60 seconds for HTC workloads. Taking it into account that the average execution time of tasks in MTC workload is only about 10 seconds, we set S as 1 second in MTC.

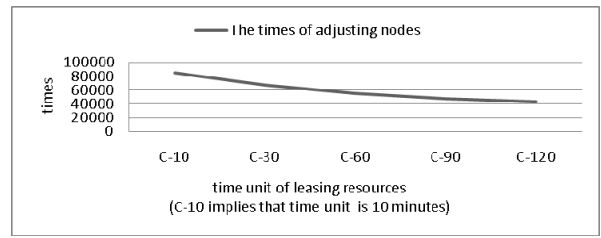

Fig.19. management overhead VS. time unit of leasing resources.

**The effect of time unit of leasing resources.**

In DawningCloud, we respectively set *the time unit of leasing resources C* as 10/30/60/90/120 minutes, while other parameters are [40B/1.5R/60S] for HTC workload and [20B/8R/1S] for MTC workload. For the resource provider, the management overhead in terms of the *accumulated times of adjusting nodes* in DawningCloud are obtained with varying *time units of leasing resources* in Fig.19. In Fig.19, *C-10* implies that *C* is 10 minutes.

From Fig.19, we have the following observation:

1) The management overhead is inversely proportional to *C*. This is because when the time unit of leasing resources is less, the service provider requests dynamic resources more frequently, which results in larger management overhead that is the sum of all service providers' times of adjusting nodes.

Taking it into account dynamic resources are charged at the granularity of *time unit of leasing resources*, we make a tradeoff and select *C* as 60 minutes in DawningCloud and the Deelman's system. In fact, in EC2 system, resources are also charged at the granularity of one hour.

**Implications of Analysis:**

At the end of this section, we give some suggestions to the service provider in setting parameters:

1) For HTC workload: the size of initial resources *B* can be set as 1/4 to 1/3 of the configuration size of the dedicated cluster system; the threshold ratio of obtaining dynamic resources *R* can be set as 1.5; the checking resource cycle *S* can be set as 60 seconds.

2) For MTC workload: *B* is set as approximate 1/8 of the configuration size of the dedicated cluster system; *R* need to satisfy the condition of *(B\*R < RA) and (RA\*R > RM)*, where *RA* is *the accumulated resource demand of jobs in queue* in most of the running time and *RM* is the *maximal accumulated resource demand of jobs in queue* in the running time (*RA* and *RM* can be calculated through experiments); *S* can be set as 1 second.

## 4.7 Total Cost Ownership of a service provider in the Evangelinos's system and the dedicated system

In this section, we compare the total cost ownership (TCO) of a service provider in Evangelinos's system and dedicated cluster systems.

For the dedicated cluster system, we take a real case from the grid lab of Beijing University of Technology, which is deployed in the year of 2006. The dedicated cluster system is composed of 15 nodes, and each node has 2*2 GHZ CPU, 4 GB memory and 160 GB DISKs; the depreciation cycle of system is 8-year; the total capital expenses (CapEx) of dedicated cluster system is $120,000. Among the operation expenses (OpEx), the total maintenance cost afforded to the company is 30,000$. The energy and space cost of the dedicated cluster system is about $1,600 per month.

For the Evangelinos's system, we choose the pricing of Amazon's EC2 Service [3] as the pricing meter. The configuration of one EC2 instance is: 2G CPU, 1.7 GB memory and 140 GB DISK; the price of the EC2 service is $0.1 per instance * hour and $0.1 per GB inbound transfer * month.

We calculate the TCO per month of the service provider in the dedicated cluster system as follows:

*TCOdcs= (CapEx depreciation) + OpEx (1)*

The TCO of the service provider in the dedicated cluster system is $3,160 per month.

We calculate the TCO per month of the service provider in EC2 as follows:

*TCO$_{ssp}$ = (Total Instance Cost) + (Inbound transfer Cost) (2)*

In order to match the configuration of the dedicated cluster system, we choose 30 EC2 instances for the service provider in EC2. The total cost of the instances is: 30day *24hours *30instances * $0.1 =$2160. From the system log, we can observe that the average data transfer per month is less than 1000 GB, so the upper cost of inbound transfer is: 1000*0.1=$100. For the Evangelinos's system (EC2), the TCO of the service provider is $2,260 per month, which is only 71.5% of that of the dedicated cluster system.

## 4.8 Analysis

From the experiment, we have two conclusions: First, from the perspectives of service providers, with respect to the dedicated cluster system, Evangelinos's system is more cost-effective, this is because a service provider has the same performance, but the TCO is less than that of the dedicated cluster system; Second, with the dynamic resource management mechanism and policies, DawningCloud outperform another two Cloud solutions: Evangelinos's system and Deelman's system from the perspectives of service providers and the resource provider.

Thus, we can conclude: *for typical workloads*, with the enabling system: DawningCloud, MTC or HTC service providers can benefit from the economies of scale on a Cloud platform.

## 5 THE PROOF OF THE ECONOMIES OF SCALE

In this section, irrespective of specific workloads, we will verify that DawningCloud can acheive the economies of scale on a Cloud platform using an analytical approach.

TABLE 5
THE PARAMETERS USED IN THE ANALYSIS.

| | |
|---|---|
| $w$ | HTC or MTC workload of a service provider |
| $D(w)$ | the time duration of $w$ |
| $c_{EM}$ | A service provider's configuration in Evangelinos's model |
| $LR(c_{EM})$ | The size of leased resources of a service provider in $c_{EM}$ |
| $c_{DC}$ | A service provider's configurations in DawningCloud |
| $IR(c_{DC})$ | the size of initial resources in $c_{DC}$ |
| $TR(c_{DC})$ | the threshold ratio of obtaining dynamic resources in $c_{DC}$ |
| $RC$ | resource consumption of a service provider |
| $PM$ | Performance of a service provider: the number of completed jobs for HTC or tasks per second for MTC. |
| $DRC(c_{DC},w)$ | For workload $w$, a service provider's dynamic resource consumption with $c_{DC}$ |
| $TRC$ | the total resource consumption of the resource provider |
| $W$ | the set of consolidated workloads of the resource provider |
| $C_{EM}$ | the set of service providers' configurations with Evangelinos's model |
| $C_{DC}$ | the set of service providers' configurations with DawningCloud |

**Lemma 1.** *For any MTC or HTC workload, there is a configuration of DawningCloud （which is represented as CONF1） that guarantees the service provider has the same resource consumption and the same performance as that of the Evangelinos's model system with any valid configuration (which is larger than the resource demand of the largest jobs).*
The formal presentation of Lemma 1 is as follows:
$\forall c_{EM}, \forall w, \exists c_{DC}$ :

$$RC(c_{DC},w) = RC(c_{EM},w) \land PM(c_{DC},w) = PM(c_{EM},w)$$

**Proof.** DawningCloud and Evangelinos's model take the same job scheduling policy.

$RC(c_{EM},w) = LR(c_{EM}) * D(w)$

$RC(c_{DC},w) = IR(c_{DC}) * D(w) + DRC(c_{DC},w)$

DawningCloud takes the resource management policy stated in Section 3.4. We set the configuration of DawningCloud CONF1 as follows: $IR(c_{DC}) = LR(c_{EM}) \land TR(c_{DC}) = +\infty$. We can get $DRC(c_{DC},w) = 0$, because under this condition, the threshold ratio of obtaining dynamic resources will not be triggered, which is also demonstrated thought experiments in Section 4.6. So $RC(c_{DC},w) = LR(c_{EM}) * D(w) = RC(c_{EM},w)$. At the same time, $PM(c_{DC},w) = PM(c_{EM},w)$ since for the same workload the service provider respectively using Evangelinos's model and DawningCloud has the same configurations.
□

**Lemma 2.** *For any MTC or HTC workload, the configuration of DawningCloud CONF1 is not always the optimal configuration that outperforms other configurations.*
The formal presentation of Lemma 2 is as follows:
$\forall w \forall c_{DC}$ : $\neg(RC(CONF1,w) \leq RC(c_{DC},w) \land PM(CONF1,w) \geq PM(c_{DC},w))$.

**Proof by contradiction.**

If
$$\forall w, \forall c_{DC} : RC(CONF1, w) \leq RC(c_{DC}, w) \land PM(CONF1, w) \geq PM(c_{DC}, w)$$

For three workload in Section 4.1, when we set $c_{DC}$ as the baseline configuration ($C_{BC}$), we get: $RC(CONF1,w) > RC(c_{BC},w) \land PM(CONF1,w) \leq PM(c_{BC},w)$ . It is a contradiction.

Hence, we can assert that the initial assumption must be false.

□

**Theorem 1.** *For any MTC or HTC workload, there is a configuration of DawningCloud that guarantees the service provider's resource consumption no more than and the performance no less than that of Evangelinos's system with any valid configuration (which is larger than the resource demand of the largest jobs).*

The formal presentation of Theorem 1 is:
$$\forall c_{EM}, \forall w, \exists c_{DC} :$$
$$RC(c_{DC}, w) \leq RC(c_{EM}, w) \land PM(c_{DC}, w) \geq PM(c_{EM}, w)$$

**Proof.** Through Lemma 1 and Lemma 2, we can get the conclusion: if *CONF1* is the optimal configuration in DawningCloud,

then: $RC(c_{DC},w) = RC(c_{EM},w) \land PM(c_{DC},w) = PM(c_{EM},w)$,

Else: $RC(c_{DC},w) < RC(c_{EM},w) \land PM(c_{DC},w) \geq PM(c_{DC},w)$.

□

**Corollary 1.** *For any sets of MTC and HTC workloads, there is a set of configurations of DawningCloud that guarantees the resource provider's total resource consumption no more than those of the Evangelinos's systems with any valid configuration (which are larger than the resource demands of the largest jobs).*

The formal presentation of Corollary 1 is:

$$\forall C_{EM}, \forall W, \exists C_{DC} : TRC(C_{DC}, W) \leq TRC(C_{EM}, W)$$

**Proof.**
$$TRC(C_{EM}, W) = \sum_W RC(c_{EM}, w),$$
$$TRC(C_{DC}, W) = \sum_W RC(c_{DC}, w).$$

FromTheorem 1,

$$\forall w, \forall c_{EM}, \exists c_{DC} : RC(c_{DC}, w) \leq RC(c_{EM}, w),$$

Hence, we have $TRC(C_{DC},W) \leq TRC(C_{EM},W)$.

□

**Discussion** From the above analysis, we have two findings: 1) For *any* MTC or HTC workload, there is a configuration of DawningCloud that enables the service provider to consume resources no more than that using Evangelinos's system with the same or better performance assurance; 2) with consolidation of MTC and HTC workloads, there is a set of configurations of DawningCloud that enable the resource provider's total resource consumption not more than that of Evangelinos's system. On the other hand, from the perspectives of service providers, with respect to the dedicated cluster system, Evangelinos's system is more cost-effective. So we can conclude: irrespective of specific workloads, MTC or HTC service providers can indeed benefit from the economies of scale on Cloud platforms with the enabling system DawningCloud in comparison with the traditional solution that small or medium scale organizations own dedicated cluster system.

# 6. RELATED WORK

In this section, we summarize the related work.

## 6.1 Evaluation of Cloud systems:

Armbrust et al. [2] in theory show the workloads of Web service applications can benefit from the economies of scale of Cloud computing systems, however, no one answers this question from the perspective of scientific communities. In the context of hybrid cloud, M. D. de Assuncao et al [30] investigate whether an organization operating its local cluster can benefit from using Cloud providers to improve the performance of its users' requests; P. Marshall et al.'s evaluation of elastic site [29] consists primarily of a comparison of the three different policies (on demand, steady stream and bursts) in an attempt to maximize job turnaround time while minimizing thrashing and idle VMs. M. R. Palankar et al [33] evaluates S3 as a black box and reasons whether S3 is an appropriate service for science grids.

## 6.2 infrastructure for scientific communities

**Public Cloud solutions**: Amazon's EC2 directly provides resources to end users, and relies upon end user's manual management of resources. EC2 extended services: RightScale (http://www.rightscale.com/) provides automated Cloud computing management systems that helps you create and deploy only *Web service applications* running on EC2 platform. There are two proposed usage models for EC2-like public clouds in scientific communities. Deelman et al. [7] propose each staff of an organization to directly lease virtual machine resources from EC2 for running applications in a specified period. Evangelinos et al. [3] propose that an organization as a whole rents resources with the fixed size from EC2 to create a leased cluster system that is deployed with a queuing system, like OpenPBS, for HTC workloads.

**Private and hybrid cloud solutions:** two open source projects, OpenNebula (www.opennebula.org/) and Haizea (http://haizea.cs.uchicago.edu/), are complementary and can be used to manage Virtual infrastructures in private/hybrid clouds [32]. In the context of hybrid cloud, recently, Sun Microsystems has added support for Amazon EC2 into Sun Grid Engine (SGE); R. Moreno-Vozmediano et al [26] analyze the deployment of generic clustered services on top of a virtualized infrastructure layer that combines a VM manager (on a local cluster) and a cloud resource provider (external cloud provider: Amazon EC2). P. Marshall et al [29] have implemented a resource manager, built on the Nimbus toolkit to dynamically and securely extend existing physical clusters into the cloud.

**Virtual execution environments:** Irwin et al. [9] [12] propose a prototype of service oriented architecture for resource providers and consumers to negotiate access to resources over time. On a basis of virtualization technologies, previous systems provide virtual execution environments either for grid computing [20] [22] [27] [28] or data centre [19] [21]. E. Walker et al [31] presents a system for creating personal clusters in user-space to support the submission and management of thousands of compute-intensive *serial jobs*, which allows the expansion of local resources on-demand during busy computation periods.

## 6.3 Resource management issues

Resource management issues are widely investigated in the context of cloud computing and grid computing. In the context of cloud computing, L. Grit et al [12] designs the Winks scheduler to support a weighted fair sharing model for a virtual Cloud computing utility. The goal of the Winks algorithm is to satisfy these requests from a resources pool in a way that preserves the fairness across flows. In the context of private cloud, B. Sotomayor et al [36] present the design of lease management architecture, Haizea, which implements leases as virtual machines (VMs) to provide leased resources with customized application environments. In the context of hybrid cloud, M. D. de Assuncao et al [30] evaluate the cost of six scheduling strategies used by an organization that operates a cluster managed by virtual machine technology and seeks to utilize resources from a remote Infrastructure as a Service (IaaS) provider to reduce the response time of its user requests. M. Dan [15] proposes the algorithm for scheduling mixed workloads in multi-grid environments, whose goal is to minimize the task's turnaround time in grid environment. In our paper, we focus on resource management issues in the context of public cloud, which is an integrated part of providing the platform helping us answering the concerned economies of scale issue.

## 7. CONCLUSION

In this paper, we have answered one key question to the success of Cloud computing: In scientific communities, can small- or medium-scale organizations benefit from the economies of scale? Our contributions are four-fold: first, we proposed a dynamic service provisioning (ESP) model in Cloud computing. In the ESP model, a resource provider can create specific runtime environments on demand for MTC or HTC service providers, while a service provider can resize dynamic resources. Second, on a basis of the ESP model, we designed and implemented an enabling system, DawningCloud, which provides automatic management for heterogeneous MTC and HTC workloads. Third, our experiments show that for typical MTC and HTC workloads, MTC and HTC service providers and the resource service provider can benefit from the economies of scale on a Cloud platform. Lastly, using an analytical approach we verify that irrespective of specific workloads, DawningCloud can achieve the economies of scale on Cloud platforms.

## ACKNOWLEDGMENT

This paper is supported by the 863 program (Grant No. 2009AA01Z128) and the NSFC programs (Grant No. 60703020 and 60933003).

## REFERENCES

[1] I. Raicu, I. Foster, and Y. Zhao. 2008. Many-Task Computing for Grids and Supercomputers. In Proceedings of MTAGS 08.

[2] M. Armbrust, A. Fox, and et al. 2009. Above the Clouds: A Berkeley View of Cloud Computing. Technical Report.

[3] C. Evangelinos, C. Hill. 2008 .Cloud computing for parallel Scientific HPC Applications: Feasibility of running Coupled Atmosphere-Ocean Climate Models on Amazon's EC2. In Proceedings of CCA08.

[4] L. Vaquero, L. Rodero-Merino. 2009. A break in the Clouds: towards a Cloud definition. ACM SIGCOMM Computer Communication Review Volume 39, Issue 1, January 2009.

[5] S. Garfinkel. 2007. Commodity Grid Computing with Amazon's S3 and EC2. Login: The USENIX Magazine, February 2007, Volume 32, Number 1.

[6] M. Palankar, A. Iamnitchi. 2008. Amazon S3 for Science Grids: a Viable Solution? In Proceedings of DADC08.

[7] E. Deelman, G. Singh. 2008. The Cost of Doing Science on the Cloud: The Montage Example. In Proceedings of ACM/IEEE SC08.

[8] D. Kondo, B. Javadi. 2009. Cost-Benefit Analysis of Cloud Computing versus Desktop Grids. In Proceeding of HCW09.

[9] D. Irwin, J. Chase. 2006. Sharing networked resources with brokered leases. In Proceedings of USENIX '06.

[10] K. Gaj, T. El-Ghazawi. 2002. Performance Evaluation of Selected Job Management Systems. In Proceedings of 16th IPDPS, pp.260-260.

[11] M. Livny, J. Basney. 1997. Mechanisms for High Throughput Computing, SPEEDUP Journal, Vol. 11, No. 1, June 1997.

[12] L. Grit, J. Chase. 2008. Weighted fair sharing for dynamic virtual clusters. In Proceedings of SIGMETRICS 2008, pp. 461-462.

[13] J. Zhan, L. Wang, B. Tu, Y. Li, P. Wang, W. Zhou, and D. Meng, 2009. Phoenix cloud: Consolidating heterogeneous workloads of large organizations on cloud computing platforms. In Proceedins of CCA 08. The modified version is available at CoRR abs/0906.1346.

[14] J. Zhan, L. Wang, W. Shi, S. Gong, X. Zang. 2010. PhoenixCloud: Provisioning Resources for Heterogeneous Cloud Workloads. arXiv :1003.0958v1.

[15] M. Dan, D. Geiger. 2006. Scheduling Mixed Workloads in Multi-grids: The Grid Execution Hierarchy. In Proceedings of 15th HPDC, pp. 291-302.

[16] J. Zhan, G. Liu, L. Wang, and et al. 2006. PhoenixG: A Unified Management Framework for Industrial Information Grid. In Proceedings of CCGrid 2006, pp. 489-496.

[17] B. Rochwerger, D. Breitgand, E. Levy, A. Galis, K. Nagin, I. M. Llorente, R. Montero, Y. Wolfsthal, E. Elmroth, J. Cáceres, M. Ben-Yehuda, W. Emmerich, and F. Galán. 2009. The Reservoir model and architecture for open federated Cloud computing，to appear, IBM J. Res. Dev., Vol. 53, No.4, 2009.

[18] M. Kallahalla, M. ysal, R. waminathan, D. E. Lowell, M. ray, T. hristian, N. dwards, C. I. alton, F. Gittler. 2004. SoftUDC: A Software-Based Data Center for Utility Computing. Computer 37, 11 (Nov. 2004), pp.38-46.

[19] P. Ruth, P. McGachey, X. Jiang, and D. Xu. 2005. VioCluster: Virtualization for dynamic computational domains. In Proceedings of Cluster 05, pp. 1-10.

[20] K. Keahey, I. Foster, T. Freeman, X. Zhang and D. Galron. 2005. Virtual Workspaces in the Grid. In Proceedings of Europar 05, pp.421-431.

[21] X. Jiang, D. Xu. 2003. SODA: a service-on-demand architecture for application service hosting utility platforms. In Proceedings of 12th HPDC, pp. 174-183.

[22] R.S. Montero, E. Huedo, I.M. Llorente, Dynamic Deployment of Custom Execution Environments in Grids. In Proceedings of ADVCOMP '08, pp.33-38

[23] B.Sotomayor, K.Keahey, I.Foster, Combining Batch Execution and Leasing Using Virtual Machines. In Proceedings of HPDC 2008.

[24] R. Buyya, C. S. Yeo, S. Venugopal, J. Broberg, and I. Brandic. 2009. Cloud computing and emerging IT platforms: Vision, hype, and reality for delivering computing as the 5th utility. Future Gener. Comput. Syst. 25, 6 (Jun. 2009), pp.599-616.

[25] L. Wang, J. Zhan, W. Shi, Y. Liang, and L. Yuan, 2009. In cloud, do MTC or HTC service providers benefit from the economies of scale? In Proceedings of MTAGS '09. ACM, New York, NY, 1-10.

[26] R. Moreno-Vozmediano, R. S. Montero, and I. M. Llorente, 2009. Elastic management of cluster-based services in the cloud. In Proceedings of ACDC '09, pp.19-24.

[27] J. S. Chase, D. E. Irwin, L. E. Grit, J. D. Moore, and S. E. Sprenkle, 2003. Dynamic Virtual Clusters in a Grid Site Manager. In *Proceedings of HPDC 03.*

[28] M. Rodríguez, D. Tapiador, J. Fontán, E. Huedo, R. S. Montero, and I. M. Llorente, 2009. Dynamic Provisioning of Virtual Clusters for Grid Computing. In Euro-Par 2008 Workshops - Parallel Processing, pp. 23-32.

[29] P. Marshall, K. Keahey, T. Freeman, 2010. Elastic Site: Using Clouds to Elastically Extend Site Resources, to appear in Prooceedings of CCGrid 2010.

[30] M. D. de Assuncao, A. di Costanzo, and R. Buyya, 2009. Evaluating the cost-benefit of using cloud computing to extend the capacity of clusters. In *Proceedings of* HPDC '09.

[31] E. Walker, J. P. Gardner, V. Litvin, and E. L. Turner. 2007. Personal Adaptive Clusters as Containers for Scientific Jobs, Cluster Computing, vol. 10(3), Sept 2007.

[32] B. Sotomayor, R. S. Montero, I. M. Llorente, and I. Foster, 2009. Virtual Infrastructure Management in Private and Hybrid Clouds. *IEEE Internet Computing* 13, 5 (Sep. 2009), pp.14-22.

[33] M. R. Palankar, A. Iamnitchi, M. Ripeanu, and S. Garfinkel, 2008. Amazon S3 for science grids: a viable solution?. In *Proceedings of* DADC '08. pp.55-64.

[34] J. Zhan, N. Sun: Fire Phoenix Cluster Operating System Kernel and its Evaluation. In Proceedings of CLUSTER' 05. Pp.1-9.

[35] C. Evangelinos, P. F. Lermusiaux, J. Xu, P. J. Haley, and C. N. Hill, 2009. Many task computing for multidisciplinary ocean sciences: real-time uncertainty prediction and data assimilation. In *Proceedings of* MTAGS '09. pp. 1-10.

[36] B.Sotomayor, R.Santiago Montero, I.Martín Llorente, I.Foster, Capacity Leasing in Cloud Systems using the OpenNebula Engine. In Proceedings of CCA08.

[37] W. Zhou, L. Wang, L. Yuan, J. Zhan, Scalable Group Management in Large-Scale Virtualized Clusters, To appear in the Journal of High Technology Letters. Available at http://arxiv.org/abs/1003.5794.